\documentstyle[preprint,aps,epsfig]{revtex}
\newcommand{\lsim}{\mathrel{\mathop{\kern 0pt \rlap
  {\raise.2ex\hbox{$<$}}}
  \lower.9ex\hbox{\kern-.190em $\sim$}}}
\newcommand{\gsim}{\mathrel{\mathop{\kern 0pt \rlap
  {\raise.2ex\hbox{$>$}}}
  \lower.9ex\hbox{\kern-.190em $\sim$}}}
\newcommand{\be}     {\begin{equation}}
\newcommand{\ee}     {\end{equation}}
\newcommand{\bea}     {\begin{eqnarray}}
\newcommand{\eea}     {\end{eqnarray}}

\begin{document}
\preprint{
\vbox{ \hbox{SNUTP\hspace*{.2em}01-038}
}}
\title{ 
Production of Spinning Black Holes at colliders
}

\author{
Seong Chan Park\footnote{schan@mulli.snu.ac.kr} 
~and ~~ H.~S.~ Song\footnote{hssong@physs.snu.ac.kr} 
}
\vspace{1.5cm}
\address{
Center for Theoretical Physics and
School of Physics, Seoul National University,
Seoul 151-742, Korea
}

\maketitle
\thispagestyle{empty}
\setcounter{page}{1}

\begin{abstract}
\noindent 
When the Planck scale is as low as TeV scale, 
there will be chances to produce Black holes (BH's)
at future colliders. 
Generally, BH's  produced via pariticle collisions
could have non-zero angular momentum.
We estimate the production cross section of 
spinning and non-spinning BH's for future colliders.
Although the production cross section for the rotating 
BH is much suppressed by angular momentum dependent factor,
the total cross section could be $\sim 2 -3$ times
enhanced for the case of $\delta =4-6$.
\end{abstract}

\vskip 0.5cm
\noindent
PACS number(s): 11.30.Pb, 11.30.Er

\newpage

{\it Introduction.--} 
Some  of the most intriguing phenomena of the TeV-scale gravity
\cite{ADD}, \cite{ADD-pheno} is  the production of Black holes(BH's)
at the future colliders like Large Hadronic Collider (LHC) of the CERN
\cite{D+L},\cite{G+T} (see also \cite{A+D+M} as earlier study on
the properties of black holes
in the large extra dimensions/TeV-scale gravity scenario)
and
at the atmosphere by cosmic rays \cite{cosmic-bh}. 
%

The mass of the BH, $M_{BH}$, produced at LHC could be larger
than the $D$-dimensional gravity scale, $M_D$, 
which is related with four dimensional Planck scale, $M_{P}$, as
$M_D^{\delta+2}R^\delta = M_P^2$,
where $D=4+\delta$ and $R$ is the size of extra dimensions.

When $M_D \sim {\cal O}(1)$ TeV, the size of the extra dimensions
is much larger than the Schwarzschild radius, $R_S$, 
of $D$-dimensional BH of mass $M_{\rm BH} \sim M_D$:
\be
R_{\rm BH} \sim \frac{1}{M_D} (\frac{M_{\rm BH}}{M_D})^{1/(\delta+1)} << R.
\ee
Since the Compton wavelength of the BH($L_{\rm BH}$) is smaller than
$R_{\rm BH}$ when $M_D<M_{\rm BH}$, the following
relation is assumed to be valid for our interests:
\be
 L_{\rm BH} < R_{\rm BH} << R.
\ee 
In Ref.\cite{D+L} and \cite{G+T}, the production cross section of
BH via particle collisions were estimated by semi-classical
arguments.  
Semi-classical approximation is strictly 
valid when $L_{\rm BH} << R_{\rm BH} $ and as $M_{\rm BH}$ approaches $M_D$
unknown stringy effects might be important.

As a first try to estimate, they just considered the case for the
non-spinning, non-charged BH production. 
It is important to note that such tiny black holes
could be generated only provided their electric charge
is zero, otherwise they would be naked singularities \cite{C+H}.
In the contrary, any object which is
produced via fusion process of colliding particles could have
angular momentum or spin. Non-spinning case could occur only when
the orbital and internal angular momenta are exactly cancelled.
We may imagine the case of exactly head-on collision of 
initial particles for the production of non-spinning BH.

The main objective of this letter is re-estimation of the 
the spinning and non-spinning BH
production cross section for the improved estimation. 
We will point out that even though the spinning BH production
cross section is suppressed, the total cross section
for the spinning and non-spinning BHs could be much
enhanced. 
That means we may have bigger chance to see the BH at
our future lab experiments.           

{\it Non-spinning BH production.--}
We first review the production process of non-spinning BHs.
Classically speaking, BH can be formed when and only when
the energy $M$ is compacted into a region whose circumference in
every direction is less than $2\pi R_{\rm BH}$. $R_{\rm BH}$ is Schwarzschild
radius of BH which is determined by mass $(M)$, angular momentum
$(J)$ and possiblly (electric) charge $(Q)$. 
From the above classical argument,
we deduce the simple picture of BH production at particle collision 
as following.
Consider two partons with the center of mass (CM) energy
$\sqrt{s} = M_{\rm BH}$ moving in opposite directions.
When the impact parameter $(b)$ is less than the Schwarzschid 
radius, the particles effectively inject the critical mass
of BH in the radius. Then a BH is formed and almost stationay
in the CM frame.
From this semi-classical arguments, we could find 
the geometrical approximation for the
cross section for producing a BH of mass $M_{BH}$ as
\be
\sigma(M_{\rm BH}) \approx \pi R_{\rm BH}^2.
\ee   
The Schwarzschild radius for the non-spinning BH is given
by \cite{M+P}
\be
R_{\rm BH}(J=0) = \frac{1}{\sqrt{\pi}M_D}
     (\frac{M_{\rm BH}}{M_D} 
     \frac{8 \Gamma(\frac{\delta+3}{2})}{\delta+2})^{\frac{1}{\delta+1}},
\ee
so the larger $M_{\rm BH}$ provides the larger $R_S$.
From the above expression for the parton level cross sections,
we can obtain the total cross section by convoluting parton distribution
function (PDF) for the initial partons:
\be
\frac{d\sigma (pp \rightarrow {\rm BH} +X)}{dM_{\rm BH}}= \frac{2 M_{\rm BH}}{s} 
\sum_{a, b} \int_{M_{\rm BH}^2/s}^1 f_a(x_a) 
f_b(\frac{M_{\rm BH}^2}{x_a s})
\hat{\sigma}(ab\rightarrow {\rm BH})|_{s=M_{\rm BH}^2}. 
\ee

There are several uncertainties in this semi-classical estimation.

Firstly, there could be quantum probability that BH is
produced even when $b > R_S$. However, as Voloshin argued
in Ref. \cite{V}, this probability might be exponentially 
suppressed by Euclidean BH action( see also Ref. \cite{C}). 

Secondly, as $M_{\rm BH}$ approaches $M_P$ (or $M_D$ for $D$-dimensional
BH), BH becomes stringy. Unfortunately we do not have enough 
information of this unknown stringy correction, yet.
In this study, we simply ignore such uncertanties.

{\it Spinning BH production. --}
Now, let us consider the case with the spinning BH.
In the Ref.\cite{M+P}, the rotating BH solution in $4+\delta$ dimensional
space-time is obtained. The rotating BH generally has inner and outer
horizons. They are described not only by $M_{\rm BH}$ but also
by the angular momentum of the BH parametrized by dimensionless
paramter $a$ like
\bea \label{rotating}
R_{\rm BH}(J) &&= 
     \frac{1}{\sqrt{\pi}M_D}
     (\frac{M_{\rm BH}}{M_D} 
     \frac{1}{1+a^2}
     \frac{8 \Gamma(\frac{\delta+3}{2})}{\delta+2})^{\frac{1}{\delta+1}}
     \nonumber \\
       &&= (\frac{1}{1+a^2})^{\frac{1}{\delta+1}} R_S (J=0),
\eea
where $a \equiv \frac{(\delta+2)J}{2 M_{BH} R_{\rm BH}}$ and $J$ is
the angular momentum of BH. 

For the produced BH from the particle collision, the angular momentum
of the BH is allowed upto $(M_{\rm BH} R_{\rm BH})$ at the 
classical limit.
The range of the
$a$ parameter is given as
\be
 0 \leq a \leq \frac{(\delta+2)}{2}.
\ee
As we will show below, the upper bound for the angular momentum
does not much affect the estimation of total cross section. 

To estimate the BH production cross section for rotating BH,
we assume that the semi-classical reasoning for the non-rotating BH
might be still valid for rotating BH.
Semi-classical reasoning suggests that, if the impact parameter
is less than the size of BH given in Eq.(\ref{rotating}), a BH
with the mass $M_{\rm BH}$ and angular momentum $J$ forms. 
The total cross section
can be estimated and is of order
\bea
\sigma(M_{\rm BH}) 
&&=\sum_J \hat{\sigma}(J) \nonumber \\ 
&&=\sum_{a=0}^{\frac{\delta+2}{2}}
                (\frac{1}{1+a^2})^{\frac{2}{\delta+1}}
                \hat{\sigma}(J=0)
\eea 
at the parton level.

We see that the production cross section for the large angular momentum
is a bit suppressed by the angular momentum dependent factor.
In Fig.1, we plotted the ratio of 
${\cal R}\equiv \hat{\sigma}(J)/\hat{\sigma}(0)$ 
with respect to $J$ (or $a$) at the parton level.
Solid(black) line and dashed(red) line describe the cases
with $\delta=4$ and $\delta=6$, respectively. 
The higher spin states of BH is seen to be quite suppressed and
we can understand this suppression from the $a$ dependent
factor $(\frac{1}{1+a^2})^{\frac{2}{\delta+1}}$. 
This factor is originally introduced to describe 
the radius of spinning BH.
With the factor the radius of the spinning BH 
is smaller than that of non-spinning
BH for the same energy and the geometrical cross section
is also suppressed since $\sigma\sim R_S^2$.  

The total cross section could be obtained after convoluting about
parton distribution fuction. With the geometrical arguments,
the spin dependent part of the BH production cross section 
could be factored out. We found that the total cross section
of spinning and non-spinning BH production process
is about $\sim 2$ and $\sim 3$ times larger than that of
only non-spinning BH production process
for $\delta=4$ and
$\delta=6$, respectively.     

{\it Decay of spinning BH. --}
Produced BH might decay primarilly via Hawking radiation.
Hence with the large extra dimension, we may have chance to 
test the Hawking's speculative BH thermodynamics \cite{H}.
There are some known properties of BH radiation for the 
BH located on the brane. 
BH acts as a point radiator and emit mostly $s$-waves 
since the wavelength of the thermal radiation of Hawking themperature
is larger than the size of the BH \cite{D+L}.
BHs radiate mainly on the brane since there are much more
particles on the brane than in the bulk \cite{E+H+M}.
Typically, TeV-scale BHs is belived to decay very fastly
($\sim 10^{-25}$ sec.)(however there are very different
estimation based on canonical and micro-canonical pictures of
BH \cite{C+H}.)

Hawking radiation is thermal process which is governed by
Hawking temperature, which is given as
\be
T_H = \frac{(\delta+1)+(\delta-1)a^2}{4\pi(1+a^2)} \frac{1}{R_{\rm BH}}.
\ee  
Since $T_H \propto 1/R_{\rm BH}$, we see that the rotating BH has
a bit larger Hawking temperature than non-rotating one. 
BH decay occurs in several stages: 
{\it `balding phase', `spin-down phase', `Schwarzschild phase',
{\rm and} `Planck phase'} \cite{G+T}. 
Through the balding phase, BH will settle down to a symmetrical
rotating BH by emitting gauge and gravitational radiation.
Then by emitting quanta with non-zero angular momenta BH will
lose its spin. In Schwarzschild phase, BH is non-rotating or 
Schwarzschild BH. After the Schwarzschild phase, the BH mass
is much reduced and the complex stringy effects  will be 
important. This is Planck phase.
In principle, we can measure the energy spectrum of the radiation
from the BH and by reconstructing Wien's displacement law,
we may have chance to test Hawking's law for rotating and
non-rotating BH.

{\it Summary and Conclusion.--}
In this letter, we study the 
spinning and non-spinning BH production at particle collider
based on the semi-classical arguments.
We point out that 
from the angular momentum dependence of the size of BH,
geometrical cross section for the rotating BH production
process is much suppressed. However the total cross section
obtained 
by summing rotating and non-rotating BH production cross sections
could be  $\sim 2 -3 $ times enhanced at the parton level.   
Rotating BH might decay by Hawking radiation
by several steps. Its temperature is a bit higher than that
of non-rotating BH with same mass.

In this study, we just consider the case when the BH is produced
on the {\it rigid} brane. However, crucial  modification 
is expected when the fluctuation of the brane is considered
\cite{brane fluctuation}, especially in the decay process of
the BH which occurs mainly on the brane.     
Still there remains stringy corrections to our estimation
and we are waiting for the further study.
{\it Acknowledgments.--}
The work was supported in part by the BK21 program and in part by
the Korea Research Foundation(KRF-2000-D00077).


\begin{figure}[tb]
\centerline{\epsfxsize=3.0truein \epsfbox{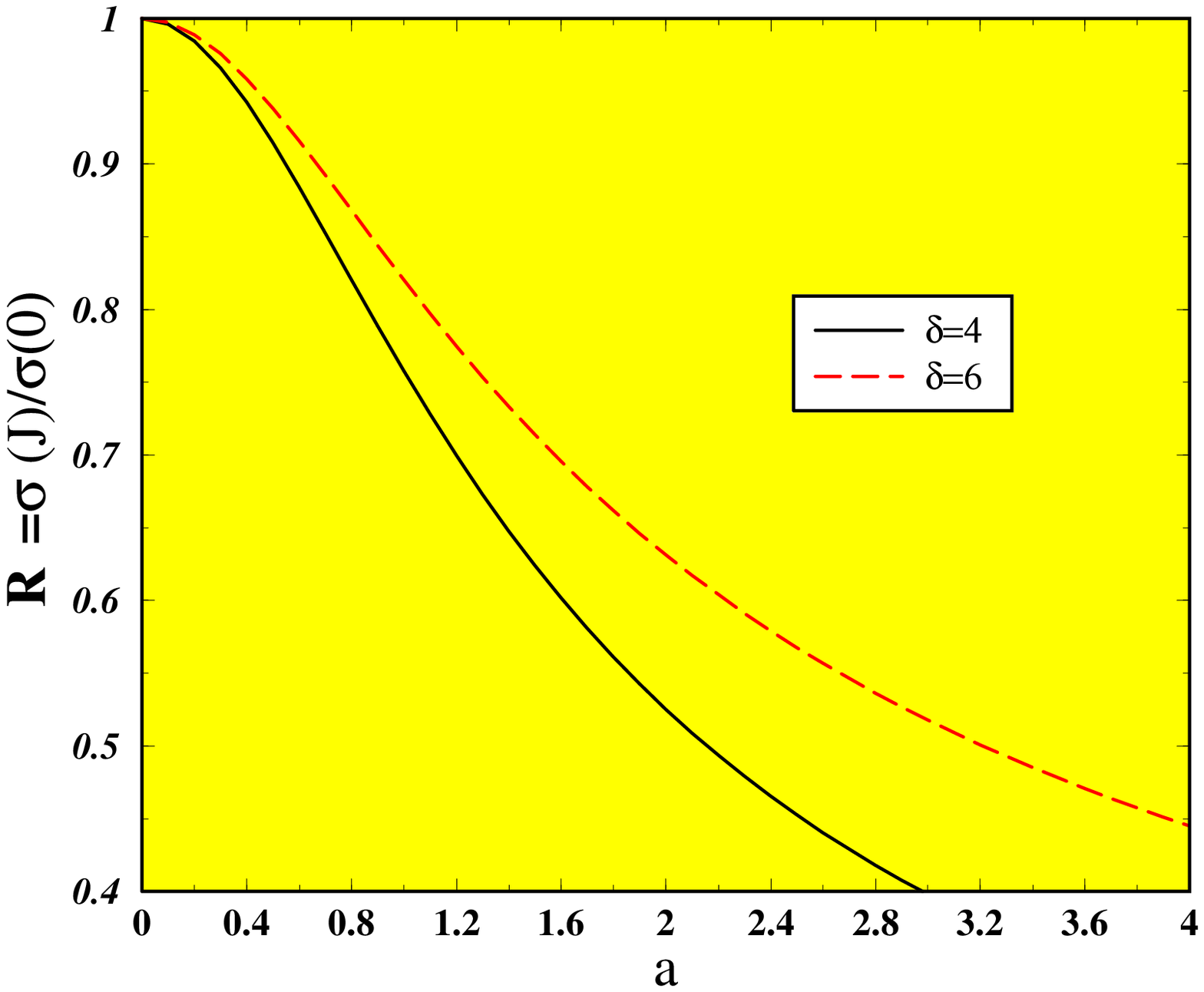}}
\caption{\it
The ratio ${\cal R}\equiv \hat{\sigma}(J)/\hat{\sigma}(0)$ plotted w.r.t
the parameter $a\equiv \frac{(\delta+2) J}{2 M_{\rm BH} R_{\rm BH}}$.
Solid(black) line and dashed(red) line
denotes $\delta=4$ and $\delta=6$, respectively.
Higher spin states for BH are quite suppressed.
}
\label{fig:ratio}
\end{figure}

\end{document}